\documentclass[journal]{IEEEtran}
\usepackage{amsmath,amsfonts}
\usepackage{algorithmic}
\usepackage{algorithm}
\usepackage{array}
\usepackage[caption=false,font=normalsize,labelfont=sf,textfont=sf]{subfig}
\usepackage{textcomp}
\usepackage{stfloats}
\usepackage{url}
\usepackage{verbatim}
\usepackage{graphicx}
\usepackage{cite}
\usepackage{hyperref}
\begin{document}

\title{Weighted cycle-based identification of influential node groups in complex networks}

\author{Wenxin Zheng, Wenfeng Shi, Tianlong Fan, Linyuan Lü,~\IEEEmembership{Senior Member,~IEEE}
        % <-this % stops a space
\thanks{This study was supported by the National Natural Science Foundation of China (Grant No. T2293771), the China Postdoctoral Science Foundation (Grant No. 2024M763131), the STI 2030—Major Projects (Grant No. 2022ZD0211400), the Postdoctoral Fellowship Program of CPSF (Grant No. GZC20241653) and the New Cornerstone Science Foundation through the XPLORER PRIZE. (Corresponding author: Tianlong Fan, Linyuan Lü.)}% <-this % stops a space
\thanks{Wenxin Zheng, Wenfeng Shi, Tianlong Fan and Linyuan Lü are with the School of Cyber Science and Technology, University of Science and Technology of China, Hefei 230026, China (e-mail: tianlong.fan@ustc.edu.cn (T.F.), linyuan.lv@ustc.edu.cn (L.L.))}}

\markboth{Journal of \LaTeX\ Class Files,~Vol.~XX, No.~XX, MARCH~2025}%
{Shell \MakeLowercase{\textit{et al.}}: A Sample Article Using IEEEtran.cls for IEEE Journals}


\maketitle

\begin{abstract}
Identifying influential node groups in complex networks is crucial for optimizing information dissemination, epidemic control, and viral marketing. However, traditional centrality-based methods often focus on individual nodes, resulting in overlapping influence zones and diminished collective effectiveness. To overcome these limitations, we propose Weighted Cycle (WCycle), a novel indicator that incorporates basic cycle structures and node behavior traits (edge weights) to comprehensively assess node importance. WCycle effectively identifies spatially dispersed and structurally diverse key node group, thereby reducing influence redundancy and enhancing network-wide propagation. Extensive experiments on six real-world networks demonstrate WCycle's superior performance compared to five benchmark methods across multiple evaluation dimensions, including influence propagation efficiency, structural differentiation, and cost-effectiveness. The findings highlight WCycle's robustness and scalability, establishing it as a promising tool for complex network analysis and practical applications requiring effective influence maximization.
\end{abstract}


\section{Introduction}
\IEEEPARstart{C}{omplex} networks have garnered significant attention as a powerful framework for abstracting and modeling diverse real-world systems, including transportation and social networks \cite{ding2019application,sun2017dominating}. A longstanding challenge in complex network research is identifying a group of nodes that exerts the most significant influence on the entire network \cite{kempe2003maximizing, li2018influence, zhou2024beyond}. Most real-world networks exhibit substantial heterogeneity and scale-free properties, where a small fraction of nodes exerts disproportionately large influence. Effectively identifying and controlling these key nodes can optimize network functionality while minimizing associated costs. This task is crucial for applications spanning disease control, rumor propagation, time-series prediction, and information dissemination \cite{hosni2020minimizing, hosni2020analysis, chen2021nursing, ren2024phase}.

Growing interest in identifying critical nodes has led to the development of numerous algorithms. These methods for detecting influential node combinations can be broadly categorized into five classes: (1) Node neighborhood-based methods, including degree centrality \cite{bonacich1972factoring,lu2016h}, semi-local centrality \cite{chen2012identifying}, and k-shell decomposition \cite{kitsak2010identification}; (2) Path-based algorithms, such as closeness centrality \cite{freeman2002centrality} and betweenness centrality \cite{freeman1977set}; (3) Eigenvector-based ranking approaches, such as eigenvector centrality \cite{ruhnau2000eigenvector} for undirected networks and PageRank \cite{brin1998anatomy} along with LeaderRank \cite{lu2011leaders} for directed networks; (4) Node contraction and removal strategies \cite{jiang2024comprehensive,peng2023unveiling}, which evaluate node importance through network disintegration experiments based on robustness metrics \cite{freitas2022graph}; and (5) Machine learning and deep learning approaches, including Least Squares Support Vector Machine (LS-SVM) \cite{wen2018fast}, infGCN \cite{zhao2020infgcn}, and CGNN \cite{zhang2022new}. While these centrality-based approaches effectively identify individual key nodes, they often struggle to select node combinations that collectively exert the greatest influence, as overlapping influence areas diminish their overall effectiveness \cite{kitsak2010identification}. Machine learning has emerged as a promising direction for addressing this challenge, but the ``black-box'' nature of many models \cite{castelvecchi2016can} limits interpretability \cite{rashid2024topological}, constraining their practical application.

Methods for identifying key node groups fall into two main categories: greedy algorithms and heuristic algorithms. Greedy algorithms iteratively select seed nodes that maximize influence spread by adding the most impactful node to the seed set in each step. Classic methods include CELF \cite{leskovec2007cost}, CELF++ \cite{goyal2011celf++}, and NewGreedy \cite{chen2009efficient}. However, these methods are computationally intensive, limiting their scalability to large networks. Heuristic algorithms, in contrast, leverage node-related information to efficiently select seed node sets. Common approaches include centrality-based methods, which rank nodes by methods like degree centrality and select the top-k nodes as spreaders. While fast, these methods can suffer from redundancy due to the ``rich-get-richer'' effect \cite{colizza2006detecting}, leading to suboptimal influence spread. In addition, researchers have proposed heuristic algorithms based on Reverse Influence Sampling (RIS)  \cite{borgs2014maximizing, tang2014influence} and community detection \cite{wang2022identifying,wang2010community}.

Recently, researchers have demonstrated that cycle structures play a crucial role in complex networks, offering a fresh perspective for evaluating the influence of nodes and edges \cite{fan2021characterizing, shi2024universal, jiang2025identifying}. Fan et al. \cite{fan2021characterizing} conducted an in-depth analysis by defining the cycle number matrix and proposed the cycle ratio to quantify node importance. Shi et al. \cite{shi2023cost, shi2024universal} introduced a cycle number indicator based on basic cycles to identify multiple influential spreaders with superior spreading capabilities and reduced initial costs. Jiang et al. \cite{jiang2023searching} proposed a novel cycle ranking approach by assessing the sensitivity of various cycles' Felder values to pinpoint important nodes. However, these methods focus solely on topological features, neglecting the behavioral traits of nodes, and are limited to unweighted networks.

To address these limitations, we propose Weighted Cycle (WCycle), a novel approach that integrates basic cycles with edge weights, capturing both network topology and node behavior traits. To validate the robustness of WCycle, we conducted experiments on six distinct real-world networks, comparing its performance against five established weighted methods. Results show that WCycle consistently selects better-ranked spreaders and achieves superior spreading efficiency. Furthermore, it demonstrates the highest level of cost-effectiveness.

To validate the robustness and effectiveness of our proposed method, WCycle, the remainder of this paper is organized as follows: Sec. \ref{sec:methods} presents the methodology, including the definition and calculation of basic cycles and the detailed formulation of WCycle. Sec. \ref{sec:data} describes the datasets used for evaluation. Sec. \ref{sec:results} provides an extensive empirical analysis, comparing WCycle with five benchmark methods across multiple dimensions. Finally, Sec. \ref{sec:results} concludes the paper, summarizing key findings and highlighting potential directions for future research.

\section{Methods}\label{sec:methods}
\subsection{Basic cycles}
In networks, the relationships between nodes exhibit complex and rich cycle structures \cite{fan2021characterizing, alvarez2021evolutionary}, which emerge from interconnected paths between nodes. Analyzing these structures helps reveal the formation and evolution of network associations while offering insights into information propagation patterns. Identifying key nodes often hinges on understanding their interactions within these cycles.

However, the vast number of cycles in networks makes it computationally infeasible to enumerate all of them, even in small networks with only hundreds of nodes \cite{fan2021characterizing,fan2019towards}. Therefore, it is essential to focus on key cycles that are computationally accessible and structurally representative. This study centers on basic cycles, a specific subset of simple cycles that forms a basis for the cycle space of a network. One critical property of basic cycles is that any cycle in the network can be represented as a linear combination of these cycles. As a result, basic cycles exhibit excellent dispersion properties, ensuring comprehensive coverage of the network's structural features.

Basic cycles can be efficiently identified from a network's spanning tree $T$. Given a network $G=(V,E)$, its cycle basis $B$ and individual basic cycles $c_k$ are formally defined as:
\begin{equation} 
B = {c_1, c_2, \ldots, c_k}, 
\end{equation}
\begin{equation} 
c_k = {(s, t) \cup P_{st}}, 
\end{equation}
where $(s, t)$ is an edge satisfying $(s, t) \in E$ and $(s, t) \notin T$. The path $P_{st}$ is the unique path in $T$ linking node $s$ to node $t$. Basic cycles provide a robust and scalable foundation for analyzing key node interactions and assessing their influence in complex networks.

\subsection{WCycle centrality}

The extent to which a node participates in basic cycles provides a strong indication of its topological importance. On the other hand, edge weights capture the behavioral traits of nodes. In real-world networks, interactions between nodes often exhibit varying intensities, which are described by edge weights. These weights reflect a node’s influence on its neighbors and are essential for characterizing node behavior.

By integrating these two features, we propose the WCycle method to comprehensively evaluate node importance. The WCycle value for a node $i$ is defined as:
\begin{equation} 
\mathrm{WCycle}_i=\frac{1}{|B|}\sum_{c_i\in B}\sum_{e\in c_i}w(e), 
\end{equation}
where $c_i$ is a basic cycle containing node $i$, $e$ is an edge in $c_i$, and $w(e)$ denotes the weight of edge $e$. The term $|B|$ represents the number of basic cycles in the network's cycle basis $B$ and serves as a normalization factor.

The WCycle value captures the extent to which a node articulates basic cycles and influences its neighbors within these cycles. Essentially, the WCycle metric encapsulates both the connectivity and local clustering characteristics of a node, while also integrating the intensity of interactions between the node and its neighbors. Nodes with higher WCycle values are typically structurally well-connected and behaviorally highly interactive, participating in more basic cycles and exerting stronger influence on their neighbors. Conversely, if a node does not participate in any basic cycle, its WCycle value is zero.

\begin{figure*}[!t]
\centering
\includegraphics[width=6.5in]{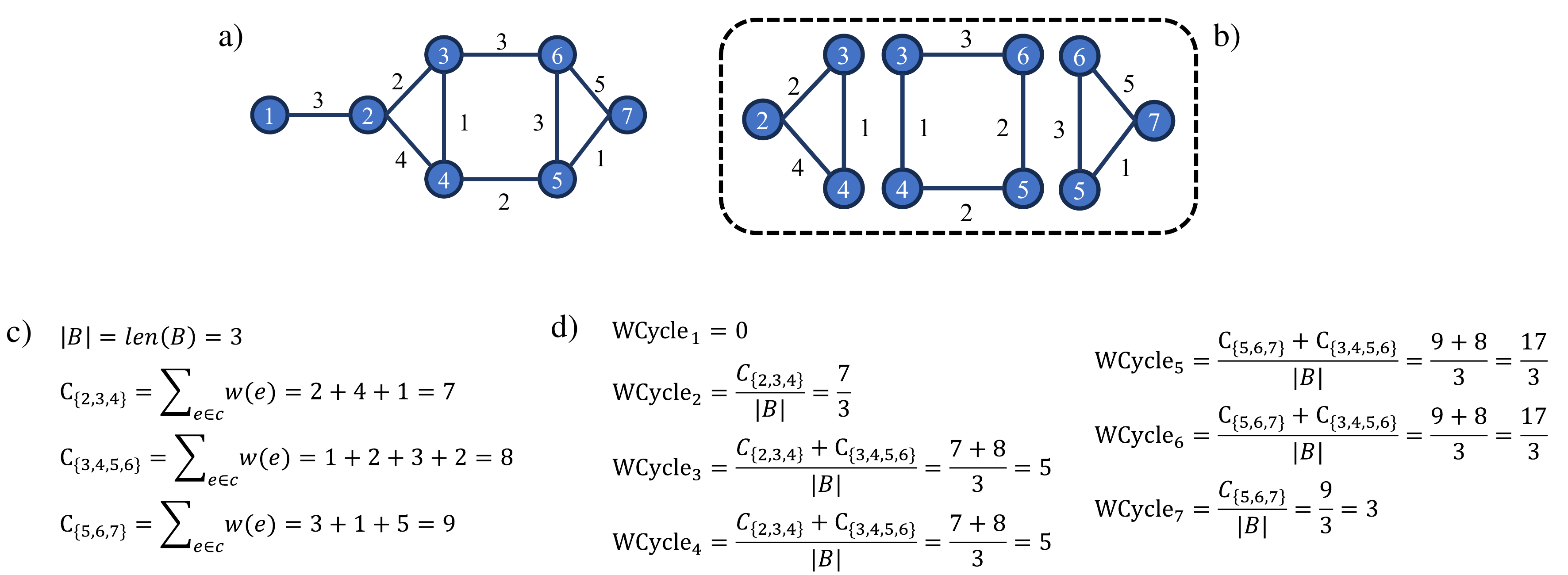}
\caption{Illustration of the WCycle centrality calculation process. \textbf{(a)} An example network $G$. \textbf{(b)} The weighted cycle basis $B$ of $G$. \textbf{(c)} Computation of cumulative edge weights for basic cycles. \textbf{(d)} Calculation of the WCycle centrality for each node.}
\label{fig_1}
\end{figure*}

The procedure for calculating the WCycle value for each node involves the following steps (see Figure \ref{fig_1}): (1) Capture the cycle basis $B$ from the network; (2) Traverse each basic cycle $c \in B$ and calculate the sum of its edge weights, $\sum_{e \in c}w(e)$; (3) Identify all basic cycles containing node $i$, sum their cumulative edge weights, and divide the result by the number of basic cycles $|B|$. Finally, based on the WCycle importance of each node, the influential node group of a network is identified by selecting the top $k\%$ nodes with the highest WCycle values.

Figure \ref{fig_1}(b) shows the weighted cycle basis of a toy network, consisting of three simple cycles, where the numbers on the edges represent their weights. Figure \ref{fig_1}(c) lists the cumulative edge weights for each weighted basic cycle, while (d) illustrates the detailed calculation process for the WCycle values of each node. The results indicate that nodes 5 and 6 have the highest WCycle values, making them the most important nodes, followed by nodes 3 and 4. Node 1 is the least important as it does not participate in any cycle, resulting in a WCycle value of zero.

\subsection{Benchmarks}

To evaluate the performance of WCycle centrality, it is necessary to select comparable centrality metrics as benchmarks for analysis. We consider five classical centrality measures and more recent approaches:

Weighted degree centrality (WD) \cite{lu2016vital}. In a weighted undirected network \(G(V,E,W)\), the weighted degree centrality (also known as strength) of a node \(v_i\) is defined as the sum of the weights of the edges connected to it:
\begin{equation}
\mathrm{WD}_i=s_i=\sum^nw_{ij}.
\label{eq.WD}
\end{equation}

Weighted H-index (WH) \cite{fan2017generalization,fan2021rise}. The weighted H-index of node \(v_i\) is computed using an \(\mathcal{H}\) function applied to a series of ordered pairs, where each pair consists of the edge weight connecting a neighbor to \(v_i\) and the neighbor's strength, as shown in Formula \eqref{eq.WH}:
\begin{equation}
\mathrm{WH}_i=\mathcal{H}[(w_{ij_1},s_{j_1}),(w_{ij_2},s_{j_2}),\ldots,(w_{ij_{k_i}},s_{j_{k_i}})].
\label{eq.WH}
\end{equation}

Weighted coreness (WC) \cite{ugander2011anatomy}. The weighted core number of node \(v_i\) is defined as:
\begin{equation}
k_i^{'}=\left[k_i^\alpha\left(\sum_j^{k_i}w_{ij}\right)^\beta\right]^{\frac{1}{\alpha+\beta}}.
\label{eq.WC}
\end{equation}
Here, \(k_i\) is the degree of node  \(v_i\), and \(\alpha\) and \(\beta\) are regulating parameters that adjust the importance of the node based on its degree and weight. In our experiments, we set \(\alpha=\beta\) to assign equal importance to both factors.

Weighted betweenness centrality (WBC) \cite{wang2008betweenness}. As a path-based centrality measure, extending betweenness centrality to weighted networks requires new definitions for shortest paths. The weighted betweenness centrality of node \(v_i\) is computed as:
\begin{equation}
\mathrm{WBC}_i=\sum_{i\neq s,i\neq t,s\neq t}\frac{g_{st}^w(i)}{g_{st}^w}.
\label{eq.WBC}
\end{equation}
where where \(g_{st}^w\) represents the total number of weighted shortest paths between nodes \(v_s\) and \(v_t\), and \(g_{st}^w(i)\) is the number of those paths that pass through node \(v_i\).

Two-way Random Walk (2RW) \cite{curado2022anew}. 2RW is an innovative betweenness centrality based on random walks. This measure evaluates a node’s significance in information dissemination by analyzing interactions among four distinct nodes while accounting for the weighted attributes of the network's edges. Due to the computational complexity of its formulation, we refer readers to the detailed definition in the literature \cite{curado2022anew}.

\section{Dataset}\label{sec:data}
To comprehensively evaluate the effectiveness and robustness of the proposed WCycle indicator, we conduct experiments on six distinct real-world networks with diverse topological and edge-weight properties:

(1) Bible Network \cite{kunegis2013konect}: This network captures the co-occurrence of nouns in the King James Version of the Bible, where edge weights represent the frequency of co-occurrences.

(2) CE-GN Network \cite{rossi2015network}: This network infers connections based on gene neighborhoods of bacterial and archaeal orthologs, with edge weights reflecting the strength of functional associations between genes.

(3) Collaboration Network \cite{newman2001structure}: This network depicts the collaboration relationships among scientists posting preprints on high-energy theory, where edge weights indicate the number of collaborations.

(4) Moreno Health Network \cite{moody2001peer}: This network represents friendships between students, with edge weights denoting the strength of their interactions.

(5) Twitter Network \cite{weng2013virality}: This network illustrates friend relationships and interactions among Twitter users, with edge weights representing the number of interaction behaviors.

(6) USA Airline Network \cite{colizza2007reaction}: This network models connections between major city airports in the USA, where edge weights correspond to the number of flights operated between airports.

\begin{table}[!t]
\caption{The Topological Properties of Real-world Networks.\label{tab:table1}}
\centering
\begin{tabular}{ccccccc} \hline  
    
        Network & $N$ &  $E$ &  $\langle k \rangle$ & $\langle w \rangle$ & $D$ & $C$  \\ \hline  
        Bible & 1707 & 9059 & 10.614 & 1.8025 & 0.0062 & 0.710  \\
        CE-GN & 2215 & 53680 & 48.470 & 1.3826 & 0.0219 & 0.184  \\ 
        Collaboration & 5835 & 13815 & 4.735 & 0.9898 & 0.0008 & 0.506\\
        Moreno\_health & 2539 & 10455 & 8.236 & 2.7149 & 0.0032& 0.147\\
        Twitter & 1996 & 16217 & 16.249 & 6.4726 & 0.0081 & 0.231 \\  
        USAir & 332 & 2136 & 12.807 & 0.0721 & 0.0386 & 0.625 \\ \hline 
    \end{tabular}
\end{table}

The structural properties of these networks are summarized in Table \ref{tab:table1}. Specifically, \(N\) and \(E\) represent the number of nodes and edges, respectively. ⟨\textit{k}⟩ denotes the average degree of nodes, while ⟨\textit{w}⟩ represents the average edge weight. Additionally, \(D\) indicates network density, and \(C\) denotes the average clustering coefficient.

\section{Results}\label{sec:results}
To comprehensively assess the effectiveness of the proposed WCycle indicator, we conduct a multi-faceted comparative analysis against five benchmark indicators across six empirical networks. The evaluation spans several dimensions, including correlations between different indicators, overlaps in identified key node groups, individuation capability, structural dispersion of identified nodes, spreading efficiency, structural similarity, and cost-performance balance. This systematic investigation aims to demonstrate the advantages of WCycle in accurately identifying influential node groups, minimizing influence redundancy, and optimizing spreading performance while maintaining cost efficiency.

\subsection{Correlations}

To assess whether Weighted Cycle (WCycle) provides unique information beyond the five benchmark indicators, we computed the Kendall’s tau (\(\tau\)) correlation \cite{knight1966computer} between all pairs of measures. The Kendall’s tau is defined as:
\begin{equation}
    \tau_b=\frac{2(N_c-N_d)}{N(N-1)},
\end{equation}
where $N_c$ is the number of concordant pairs, and $N_d$ is the number of discordant pairs in a two-by-two comparison. A correlation of 0 indicates no association between the two indicators. Values closer to 1 suggest highly similar rankings, while those closer to -1 indicate completely inverse rankings.

\begin{figure}[!t]
\centering
\includegraphics[width=3in]{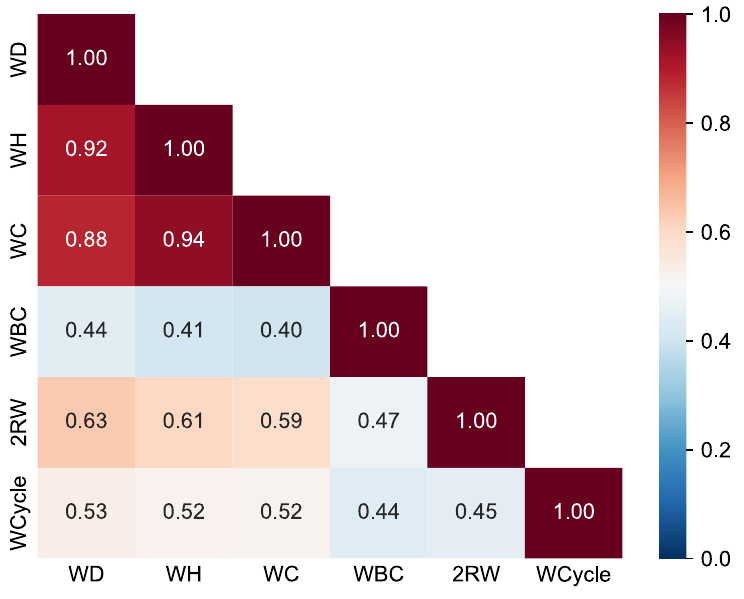}
\caption{The average correlation matrix for the six indices of node importance over six real-world networks.}
\label{fig_2}
\end{figure}

Figure \ref{fig_2} presents the average correlation matrix for the six indicators across six real-world networks. WD, WH, and WC exhibit strong correlations, indicating highly similar rankings due to their reliance on degree- and weight-based local centrality, which ranks nodes based on similar topological features. Conversely, WBC, as a global metric, shows the lowest similarity with the other indicators. WCycle, however, displays intermediate correlations, as it captures topological and behavioral information often involving nodes beyond immediate neighbors. This moderate correlation suggests that WCycle maintains a balanced association with the other indicators, offering a distinct perspective in network analysis.

\subsection{Node overlap analysis}\label{sec:overlap}

To assess the uniqueness of the important node groups identified by each indicator, we adopt Jaccard similarity, defined as:
\begin{equation} 
J(i_1,i_2)=\frac{|N_1\cap N_2|}{|N_1\cup N_2|}, 
\end{equation}
where  \(N_1\) and \(N_2\) represent the important node sets identified by indicators \(i_1\) and \(i_2\) respectively. Jaccard similarity measures the proportion of overlapping nodes relative to the union of the two sets. A lower Jaccard similarity suggests that the identified nodes are distinct, implying that WCycle captures novel information compared to the benchmark indicators.

To intuitively present the results, we compute the average Jaccard similarity for WCycle with respect to the other $m=5$ methods as follows:

% \begin{equation} J_\mathrm{WCycle}=\frac{J(\mathrm{WCycle},\mathrm{WD})+J(\mathrm{WCycle},\mathrm{WH})+\cdots+J(\mathrm{WCycle},\mathrm{2RW})}{m}. 
% \end{equation}
{\small
\begin{align}
J_\mathrm{WCycle} &= \frac{J(\mathrm{WCycle}, \mathrm{WD}) + J(\mathrm{WCycle}, \mathrm{WH}) + \cdots}{m} \nonumber \\
&\quad \frac{\cdots + J(\mathrm{WCycle}, \mathrm{2RW})}{m}.
\end{align}
}

Tables \ref{tab:table2} presents the average Jaccard similarity values for the top 5\% of nodes selected by each indicator across six empirical networks. Bold values highlight the indicator with the lowest similarity, indicating more distinct node selections.

The result shows that WCycle consistently achieves the lowest average Jaccard similarities across all networks, demonstrating its ability to select node groups that exhibit minimal overlap with those identified by other indicators. We also examined other parameters (top-3\% and 10\%), and the results remained the same. This suggests that WCycle provides a distinct analytical perspective by capturing network information that differs from the common features emphasized by both local and global centrality-based indicators.

\begin{table}[!t]
\setlength{\tabcolsep}{5pt}
    \caption{Average Jaccard Similarity for the Top 5\% of Nodes Across the Six Networks.\label{tab:table2}}
    \centering
    \begin{tabular}{ccccccc} \hline     
Networks&	$J_\mathrm{WD}$	& $J_\mathrm{WH}$ &	$J_\mathrm{WC}$	& $J_\mathrm{WBC}$ & $J_\mathrm{2RW}$	& $J_\mathrm{WCycle}$ \\ \hline
        Bible & 0.5225 & 0.5071 & 0.4698 & 0.3301 &0.4496 & \textbf{0.0206}  \\ 
        CE-GN & 0.2975 & 0.2312 & 0.0899 & 0.1943 &0.2645 & \textbf{0.0141} \\ 
        Collaboration & 0.3630 & 0.3967 & 0.3531 & 0.2121 &0.3066 & \textbf{0.0537}  \\ 
        Moreno\_health & 0.2402 & 0.2578 & 0.1862 & 0.0428 &0.2164 & \textbf{0.0157}  \\ 
        Twitter & 0.3107 & 0.3237 & 0.2883 & 0.0580 &0.1323 & \textbf{0.0112}  \\ 
        USAir & 0.5593 & 0.5166 & 0.3454 & 0.3796&0.5116 & \textbf{0.2080}  \\ \hline
    \end{tabular}
\end{table}

\subsection{Individuation analysis}

To evaluate the capacity of WCycle to distinguish important nodes, we analyze its individuation capability, which measures the ability of an indicator to assign unique importance scores to nodes. This is crucial because many traditional indicators often assign identical scores to multiple nodes, leading to insufficient differentiation. The Individuation measure $\gamma(i)$ \cite{wen2020vital} is defined as:
\begin{equation}
    \gamma(i)=\frac{N_{S}(i)}{|N|},
\end{equation}
where \(N_{S}(i)\) is the number of nodes with a unique score assigned by index $i$, and \(|N|\) is the total number of nodes in the network. A higher individuation value implies stronger node differentiation. Given the varying sizes of the networks, we adapt \(|N|\) to focus on the top-ranked node groups. Specifically, for large networks such as Bible, CE-GN, Collaboration, and Twitter, we consider the top 30\% of nodes. For the smaller network USAir, the top 50\% is selected.

\begin{table}[!t]
\setlength{\tabcolsep}{1.9pt}
\caption{The Individuation \(\gamma\) (·) of the Six Methods in the Real Networks.\label{tab:table3}}       
\centering
\begin{tabular}{ccccccc} \hline     
Networks&	$\gamma(\mathrm{WD})$	& $\gamma(\mathrm{WH})$ &	$\gamma(\mathrm{WC})$	& $\gamma(\mathrm{WBC})$ & $\gamma(\mathrm{2RW})$	& $\gamma(\mathrm{WCycle})$\\ \hline
Bible&	0.2242&	0.1676&	0.1053&	0.9844 &0.3782 & \textbf{0.9942} \\
CE-GN&	\textbf{0.9970} &	0.8872&	0.0707&	0.9865 &0.8857	& \textbf{0.9970} \\
Collaboration&	0.2210&	0.3010&	0.1576&	0.9046 & 0.0697 &	\textbf{0.9863} \\
Moreno\_health&	0.0669&	0.0367&	0.0091&	\textbf{1.0000} & 0.1155&	0.9974\\
Twitter&	0.4958&	0.3656&	0.1953&	\textbf{1.0000} &0.5092	& 0.9716 \\
USAir&	0.9939 &	0.9277&	0.6767&	0.9578&0.3855	& \textbf{1.0000} \\ \hline
\end{tabular}
\end{table}

Table \ref{tab:table3} presents the individuation values of the six methods across the real-world networks. The bold values indicate the highest individuation for each network. The results reveal that WCycle consistently achieves the highest individuation values across most networks, highlighting its superior capability to differentiate important nodes. This ability makes WCycle a reliable tool for identifying unique sources of influence, particularly in scenarios where differentiation among high-ranking nodes is critical.
\begin{figure*}[!t]
\centering
\includegraphics[width=6in]{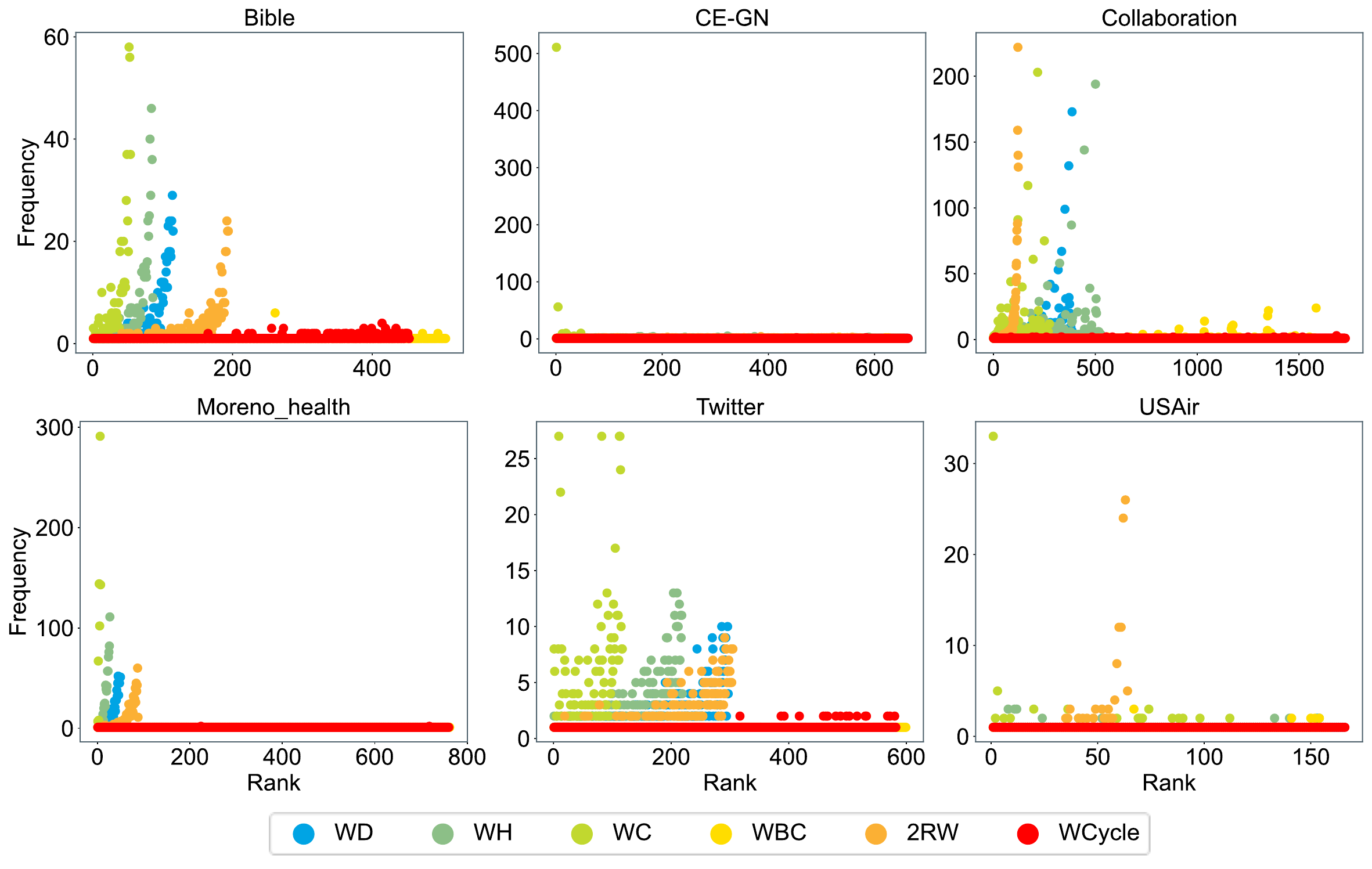}
\caption{Frequency distribution of the same ranking for the six indicators over the real-world networks.}
\label{fig_3} 
\end{figure*}

Figure \ref{fig_3} illustrates the frequency distribution of the same rankings for WCycle and the other five benchmarks. The $x$-axis represents the rankings of top nodes, while the $y$-axis shows the frequency of shared scores. Nodes sharing identical scores are assigned the same rank. Visually, WCycle exhibits the lowest frequency of shared rankings, indicating a stronger capacity for differentiation compared to other methods. In contrast, WH and WC exhibit higher frequencies of shared rankings, suggesting weaker differentiation capabilities. WD and 2RW show variable performance across networks, lacking consistency in their ability to distinguish key nodes. These findings underscore WCycle's advantage in providing fine-grained differentiation of important nodes.

\subsection{Dispersion analysis}\label{sec:Dispersion}

\begin{figure}[htbp]
\centering
\includegraphics[width=3in]{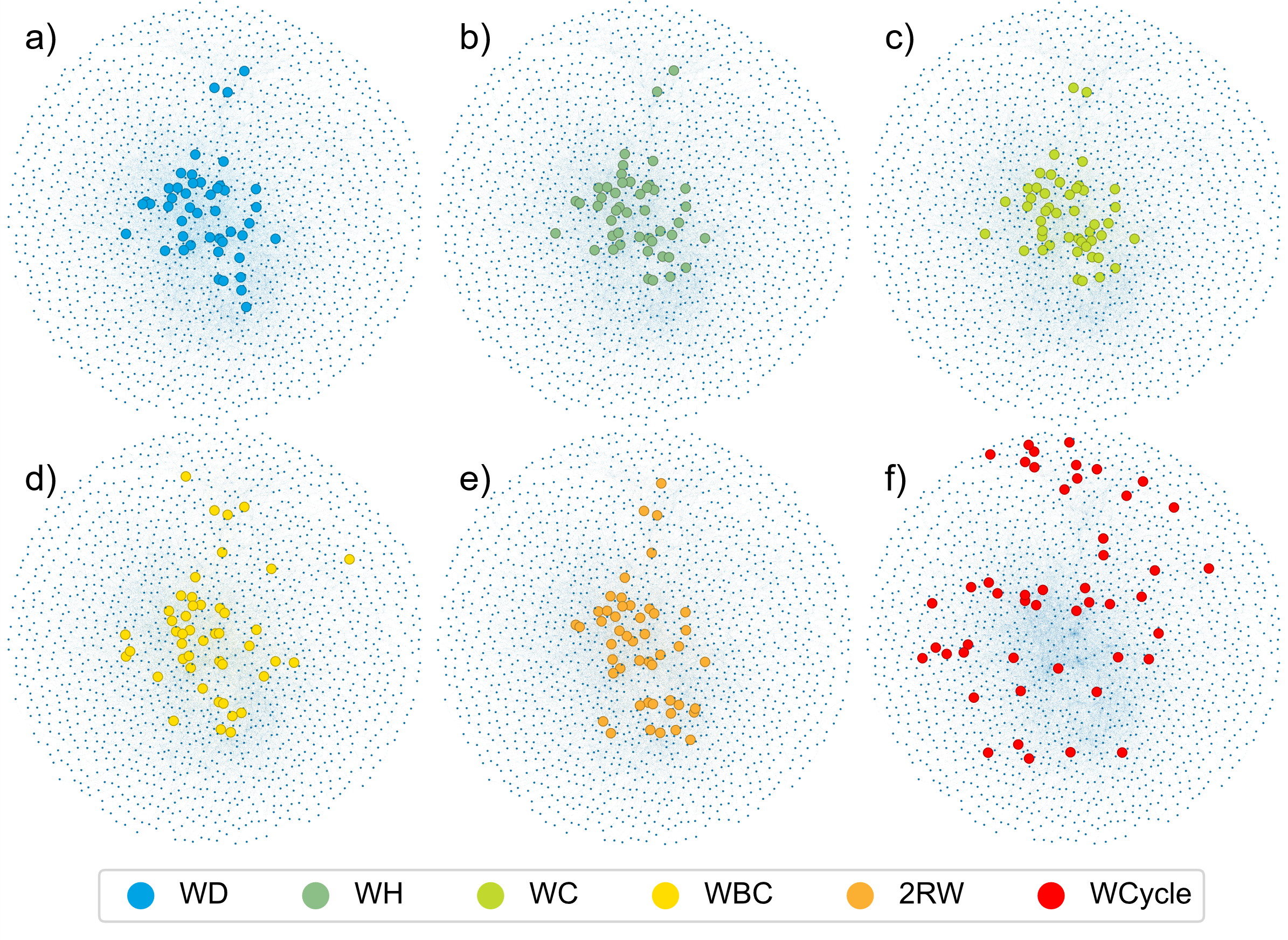}
\caption{Top-50 nodes identified by six indicators in Bible networks. (a) WD; (b) WH; (c) WC; (d) WBC; (e)2RW; (f) WCycle}
\label{fig_4}
\end{figure}

 To evaluate the dispersion of key node groups identified by WCycle and five benchmark indicators, this section presents a visual analysis of the top-50 important nodes selected by each indicator in the Bible network. Figure \ref{fig_4} illustrates the identified nodes, where node importance is represented by colors (see figure \ref{fig_4} legend) and size (larger for higher importance).

The results show that the top-50 important nodes identified by classical degree- and weight-based indicators (WD, WH, and WC) tend to form interconnected groups concentrated in specific regions of the network, demonstrating the ``rich club'' effect \cite{colizza2006detecting}. Similarly, WBC and 2RW select nodes that are centrally located, which limits their ability to fully cover diverse network areas. In contrast, WCycle selects key node groups that are better spatially dispersed across the network, with fewer interconnections. This suggests that WCycle is more effective at minimizing the overlap of influence among selected nodes, which is critical for maximizing the overall network influence. We also examined other situation (top-20), and the conclusions remained the same.

To quantitatively assess node dispersion, we calculate the average shortest path length ($d_c$) between nodes in each key node group using the following formula:
\begin{equation}\label{formula:d_c}
d_c=\frac{\sum_{i=1}^{c} \sum_{j=1, j\neq i}^{c} d_{ij}}{c(c-1)}, 
\end{equation}
where $d_{ij}$ denotes the shortest path length between nodes ranked $i$ and $j$, $c$ represents the number of selected top-$c$ important nodes. A larger value of $d_c$ indicates greater spatial dispersion within the key node group.

\begin{figure*}[!t]
\centering
\includegraphics[width=5in]{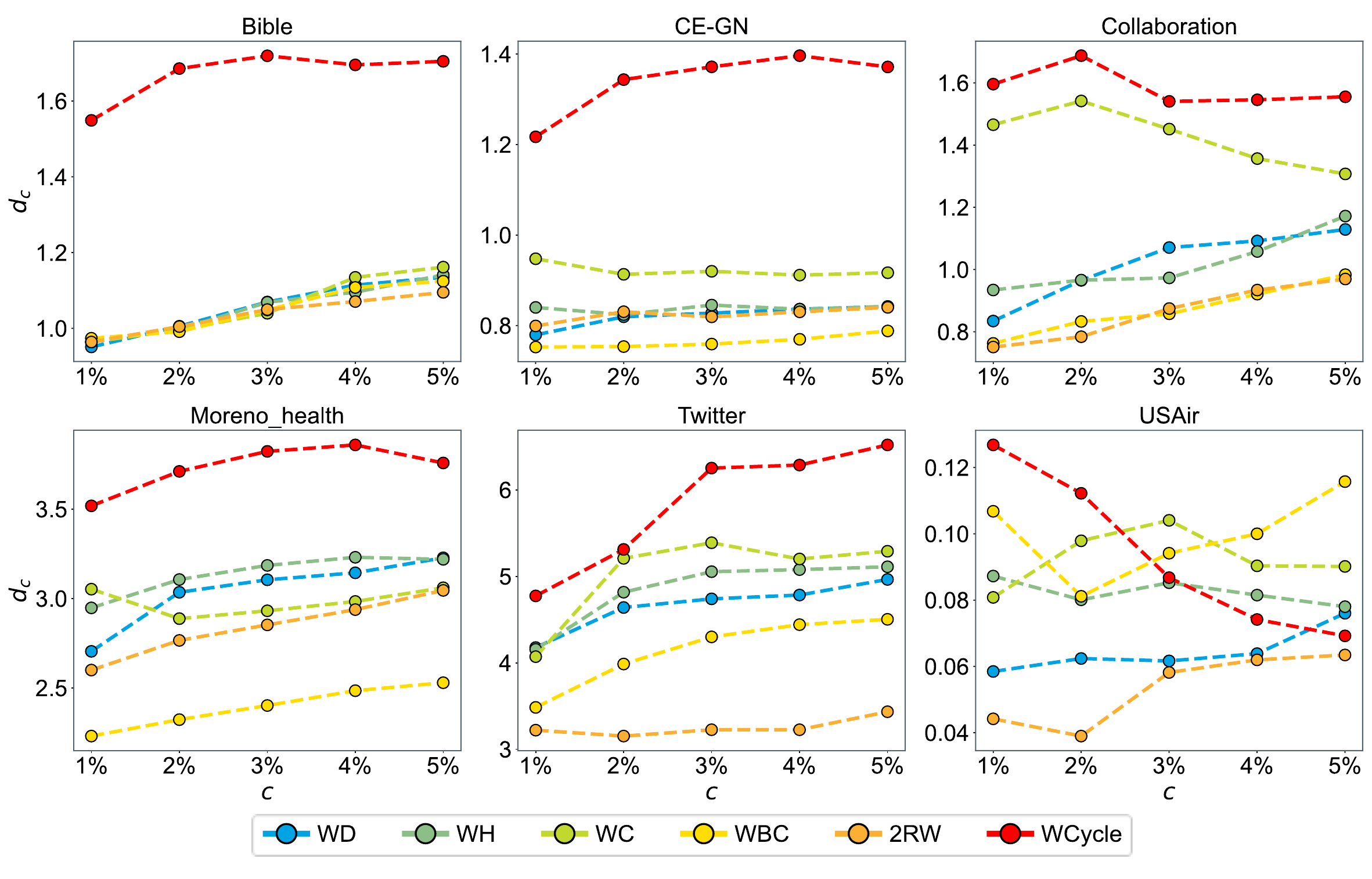}
\caption{The variation of the average shortest path length $d_c$ with different values of $c$ for the six indicators across six networks.  }
\label{fig_5}
\end{figure*}

Figure \ref{fig_5} illustrates how \(d_c\) changes with different values of \(c\) for each indicator across six networks. The results show that WCycle achieves the highest \(d_c\) values for all selected \(c\) values in five networks, indicating that the key node groups it identifies are more spatially dispersed compared to other indicators. In the USAir network, WCycle still outperforms all other indicators at \(c = 1\%\) and \(2\%\). These findings highlight WCycle's overall advantage in selecting more dispersed key node groups, reducing influence overlap and enhancing the potential for effective information propagation across most networks.

\subsection{Spreading performance analysis}

After examining various advantages of WCycle, we now focus on evaluating its core performance in maximizing influence propagation using the Weighted Susceptible-Infected-Recovered (WSIR) model \cite{sun2014epidemic}. This model extends the traditional SIR framework by incorporating edge weights. Specifically, it adjusts the propagation probability based on the weight of connections between nodes and defines the infection threshold by considering the average edge weight across the network:

\begin{equation}
    \beta_c=\frac{\langle k \rangle}{\langle w \rangle (\langle k^{2} \rangle - \langle k \rangle)},
\end{equation}
where \(\langle k \rangle\) and \(\langle k^{2} \rangle\) denote the average degree and degree squared mean, respectively, and \(\langle w \rangle\) is the average edge weight. The key objective is to select the top c\% important nodes as the seed node groups and compute the cumulative number of infected nodes, $R$, to evaluate their spreading influence.

To ensure robust experimental results, we set the recovery rate \(\mu=0.5\) and defined the infection threshold as \(\beta=\beta_c\). Each experiment was repeated 300 times under same configurations, and the average number of infected nodes was recorded as the measure of spreading influence.

Two experimental scenarios were designed to evaluate the effects of different factors on propagation performance. First, we examined the impact of varying the proportion of source nodes $c$ while keeping the infection rate fixed at \(\beta = 1.5\beta_c\). The values of $c$ were set to 1\%, 2\%, 3\%, 4\%, and 5\%. Second, we explored the effect of varying infection rates while fixing the source proportion at $c=3\%$. The infection rates were set to \(\beta_c\), 1.5\(\beta_c\), 2\(\beta_c\), 2.5\(\beta_c\) and 3\(\beta_c\).

\begin{figure*}[!t]
\centering
\includegraphics[width=5in]{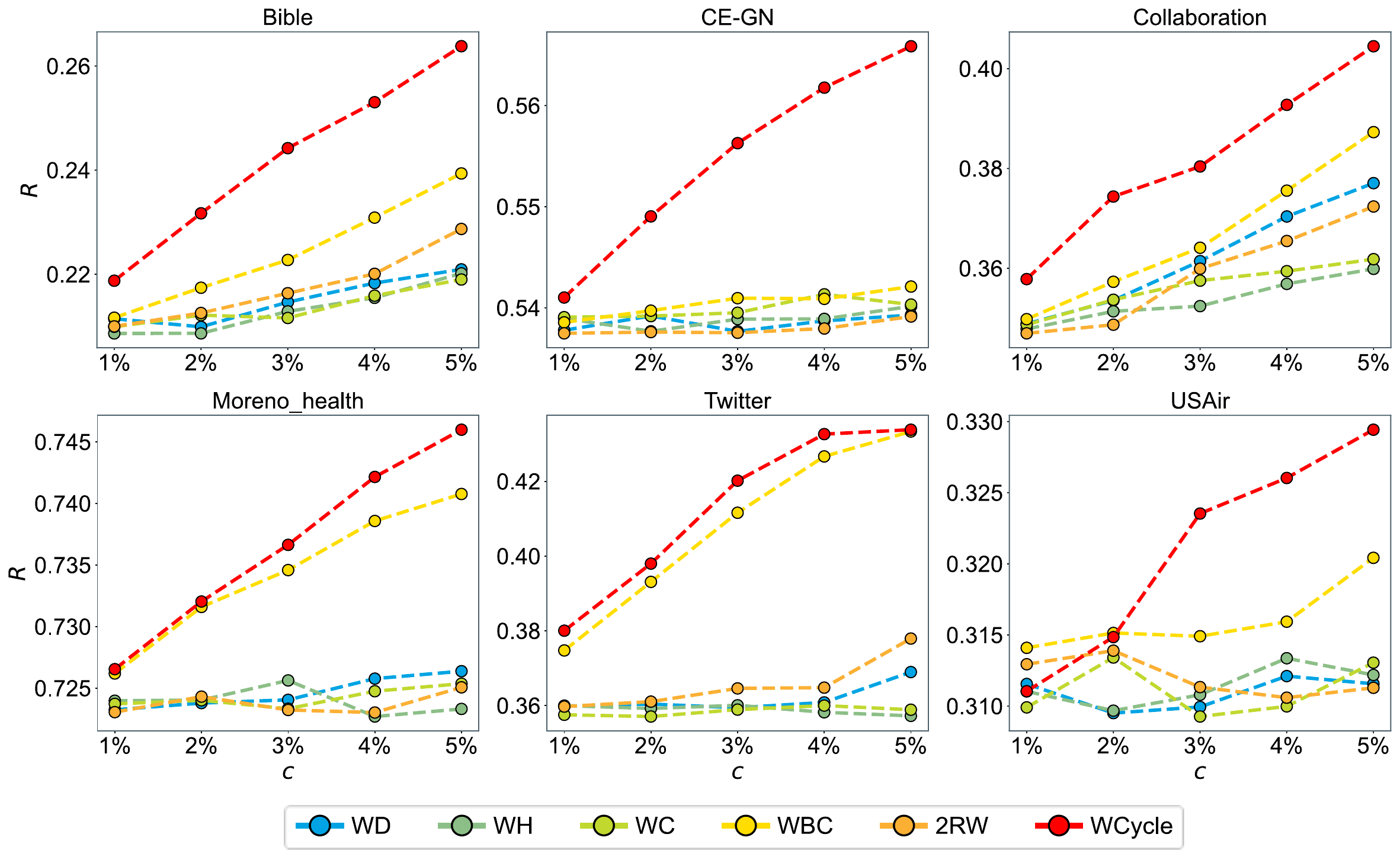}
\caption{Spreading performance of each indicator under varying source proportions at a fixed infection rate \(\beta\)=\(\beta_c\).}
\label{fig_6}
\end{figure*}

Figure \ref{fig_6} shows the influence propagation capacity of each indicator under different source proportions. WCycle maintains the highest propagation capacity in most scenarios, demonstrating superior spreading performance. In the USAir network, WCycle initially ranks second at $c=1\%$ and 2\% but surpasses other benchmarks when $c$ exceeds 2\%. This indicates that WCycle can sustain its influence as the source proportion increases. Notably, as the source proportion grows, several benchmark indicators exhibit stagnation or even a decline in propagation capacity. This phenomenon is likely due to the ``rich club'' effect, where densely connected nodes create overlapping areas of influence, limiting further propagation. WCycle effectively mitigates this limitation by selecting better-dispersed key node groups, leading to enhanced influence.

\begin{figure*}[!t]
\centering
\includegraphics[width=6.5in]{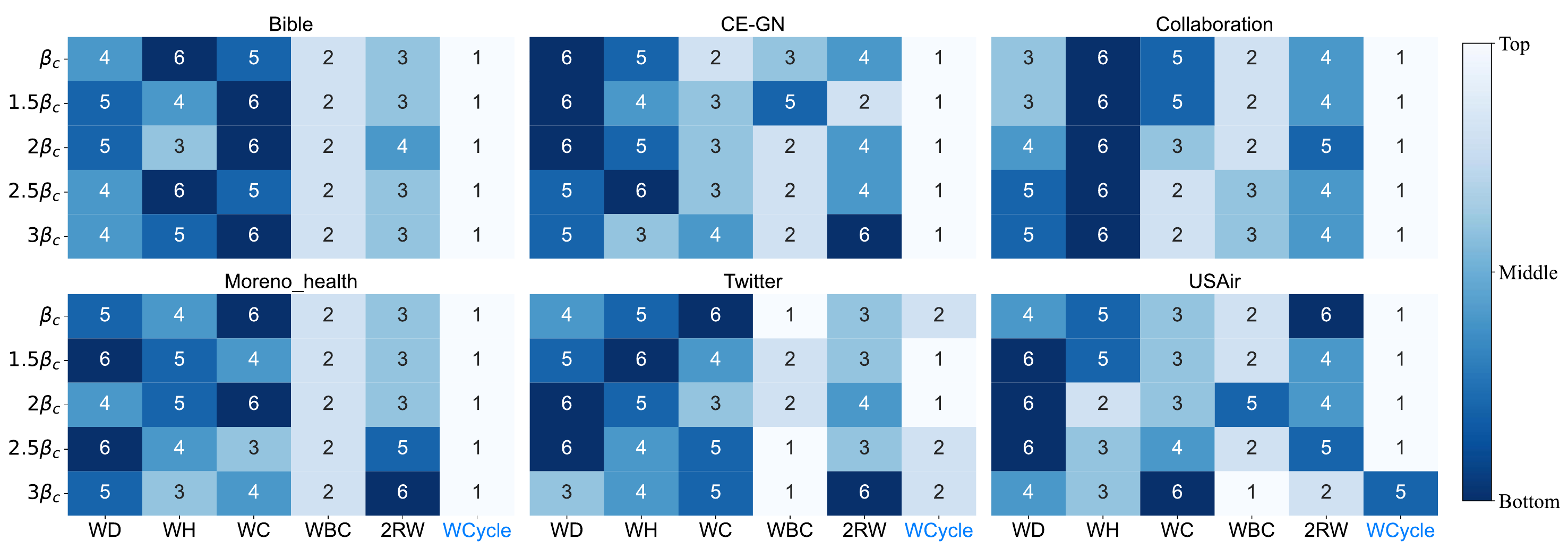}
\caption{Spreading performance of each indicator at a fixed source proportion ($c=3\%$) under different infection rates.}
\label{fig_7}
\end{figure*}

Figure \ref{fig_7} illustrates the propagation performance under varying infection rates. WCycle consistently outperforms other methods across all networks. In large-scale networks like Bible and Collaboration, it demonstrates a significant advantage. In smaller networks like USAir, while its advantage is less pronounced, it still maintains the leading position.

These findings demonstrate that WCycle achieves consistently high propagation performance across different source proportions and infection rates. Its ability to identify well-dispersed key node groups minimizes influence overlap and maximizes network-wide impact, making it a robust and effective method for influence maximization.

\subsection{Structural similarity analysis}
To further investigate why WCycle demonstrates superior spreading performance, we analyze the structural differences among the important nodes identified by each indicator. While previous analyses focused on correlations, overlaps, and the spatial dispersion of selected node groups at the set level, this section shifts the focus to differences at the node level, specifically examining their structural similarity.

We adopt Jaccard similarity \cite{jaccard1912distribution} to quantify the structural similarity between pairs of important nodes, defined as:
\begin{equation} 
J_{ij}=\frac{|\Gamma(i)\cap\Gamma(j)|}{|\Gamma(i)\cup\Gamma(j)|},
\end{equation}
where $\Gamma(i)$ represents the neighbor set of node $i$, and \(J_{ij}\) denotes the structural similarity between nodes ranked \(i\) and \(j\). A higher \(J_{ij}\) implies greater overlap in their neighborhoods.

To evaluate the overall structural similarity of important nodes selected by each indicator, we define the average structural similarity $J_{c}$ as:
\begin{equation} 
J_{c}=\frac{\sum_{i=1}^c \sum_{j=1, j\neq i}^c J_{ij}}{c(c-1)}, 
\end{equation}
where $c$ is the number of selected source nodes. A lower $J_{c}$ indicates less overlap among neighborhoods, suggesting greater topological diversity among the selected important nodes.

\begin{figure*}[!t]
\centering
\includegraphics[width=5in]{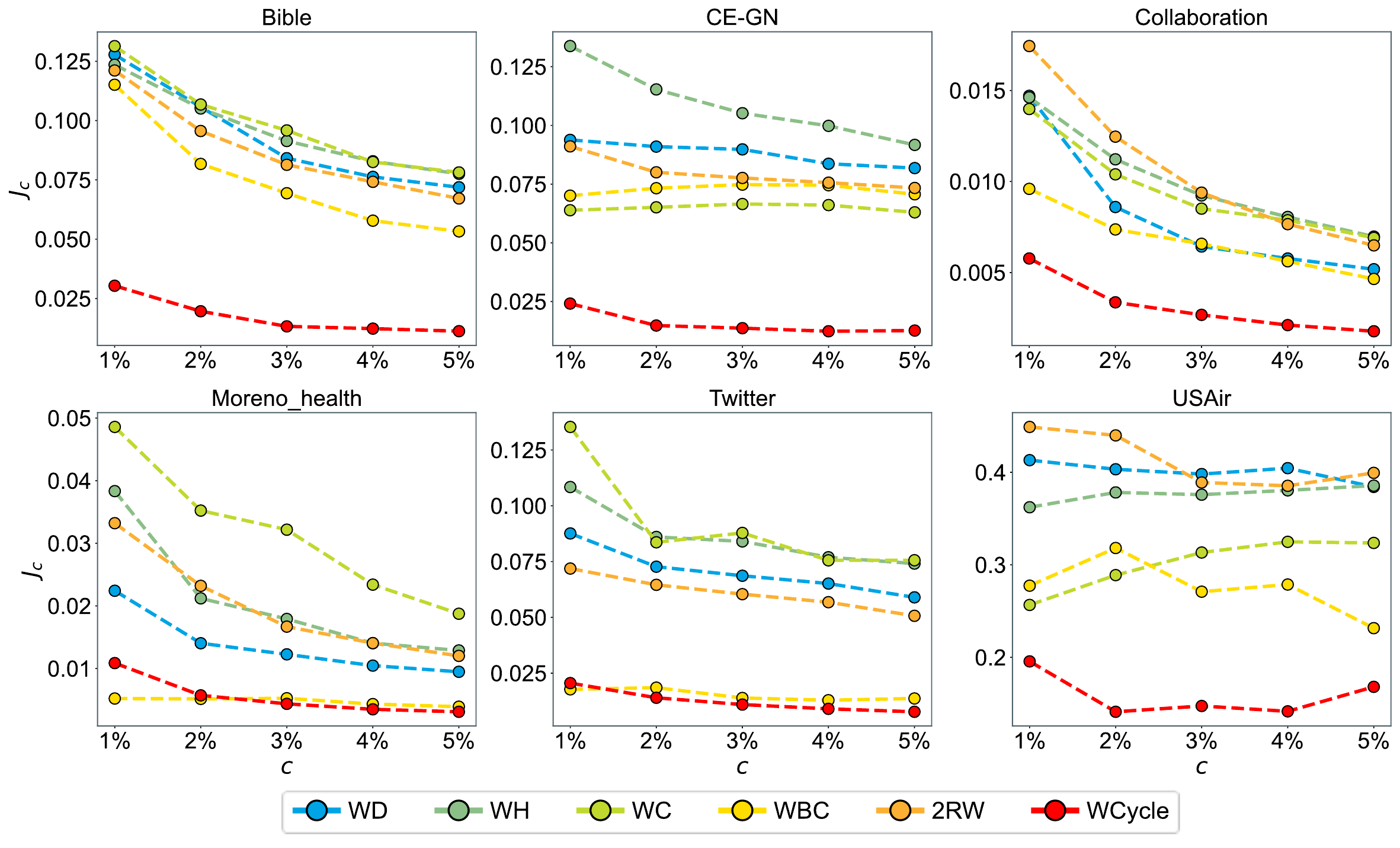}
\caption{Comparison of the average structural similarity $J_{c}$ of the top-$c$ source nodes selected by the six indicators in the six empirical networks.}
\label{fig_8}
\end{figure*}

Figure \ref{fig_8} presents the average structural similarity of the top-$c$ source nodes identified by the six indicators across six empirical networks. WCycle consistently achieves the lowest $J_{c}$ values across different values of $c$, with particularly significant differences in networks such as Bible, CE-GN, Collaboration, and USAir. This indicates that the important nodes identified by WCycle exhibit the most distinct topological features, compared to those identified by other indicators.

\begin{figure*}[!t]
\centering
\includegraphics[width=5in]{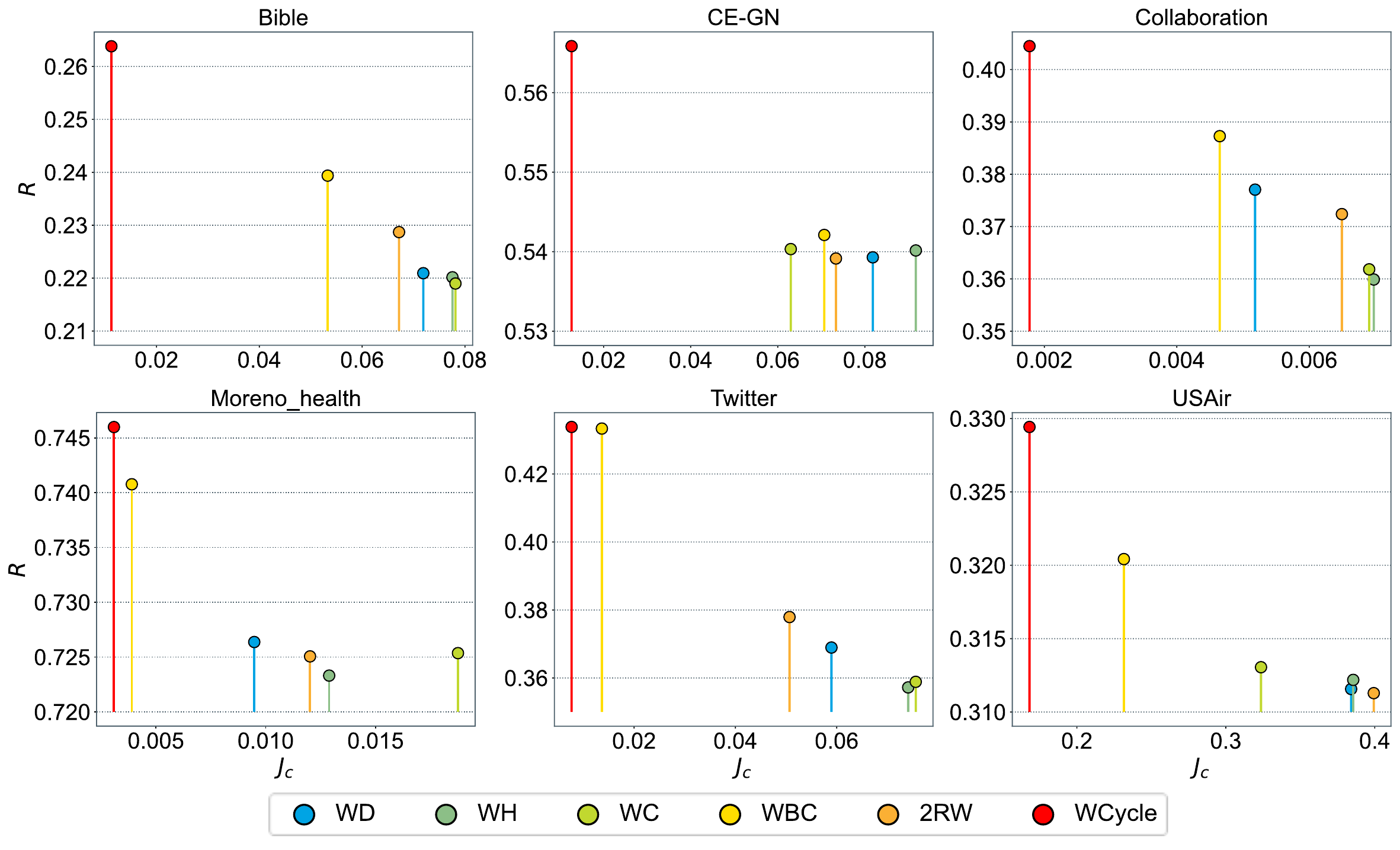}
\caption{The relationship between the average structural similarity ($J_{c}$) and the spreading capability ($R$).}
\label{fig_9}
\end{figure*}

Figure \ref{fig_9} further illustrates the relationship between the $J_{c}$ and the spreading capability $R$ of the top 5\% source nodes. An inverse correlation is evident—higher structural similarity tends to correspond to lower spreading capabilities. Among the six indicators, WCycle consistently achieves the smallest $J_{c}$ and the highest $R$, highlighting its ability to select structurally diverse important nodes.

The structural similarity analysis reveals that WCycle identifies important nodes with greater topological diversity compared to other indicators. This differentiation reduces influence redundancy and contributes to its superior spreading performance. The findings underscore the importance of selecting structurally distinct nodes in maximizing network influence.

\subsection{Cost-effectiveness analysis}
In real-world scenarios, activating important source nodes often involves varying costs, which depend on factors such as the influence exerted by the nodes and their availability within the network. Therefore, a critical challenge is achieving the desired dissemination effect at the lowest possible cost.

Previous studies \cite{ji2017effective} suggest that the cost of activating spreaders depends on two key factors: influence and scarcity. Influence is traditionally measured by node degree \(k\), while scarcity is quantified by the probability \(p(k)\) of finding nodes with degree \(k\) in the network. As \(k\) increases, \(p(k)\) decreases, leading to higher corresponding costs. 

In weighted networks, we adapt this approach by defining influence as node strength \(s\) and scarcity as the probability \(p(s)\). The total cost \(\lambda\) for activating the top $c$ spreaders is computed as:
\begin{equation} 
\lambda=\sum_{i=1}^c\frac{s}{p(s)}.
\end{equation}

\begin{figure*}[!t]
\centering
\includegraphics[width=5in]{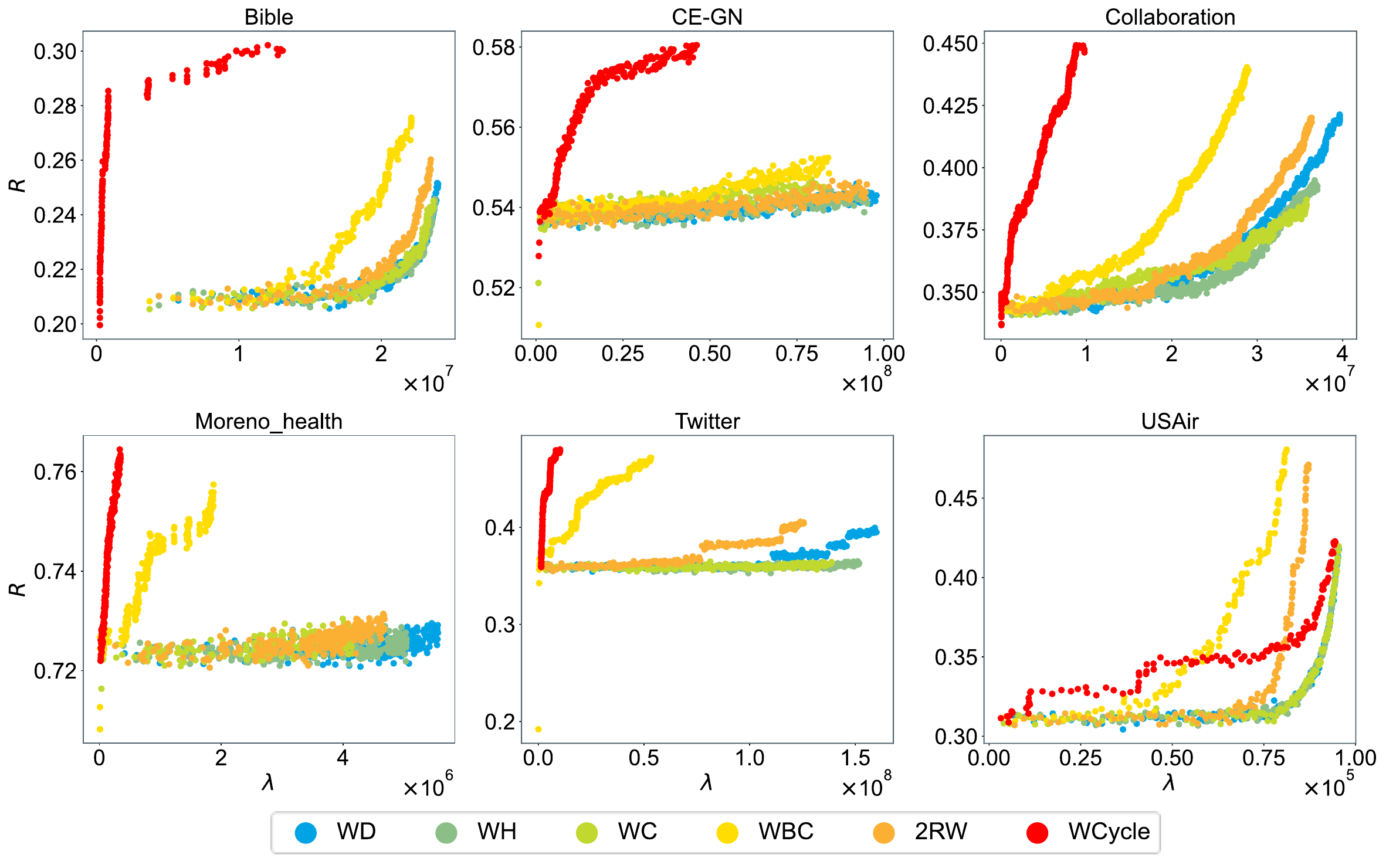}
\caption{Spreading capability ($R$) versus initial cost ($\lambda$) for the six indicators.}
\label{fig_10}
\end{figure*}

Figure \ref{fig_10} illustrates the relationship between the total cost ($\lambda$) and the overall spreading capability ($R$) for top-$c$ nodes, where $c$ ranges from 2 to 10\% of the network size. The results indicate that WCycle consistently achieves the highest spreading capability for a given cost across five networks, demonstrating a distinct cost-performance advantage. Even in the USAir network, WCycle maintains a competitive position during the early stages. Notably, a slight increase in WCycle's cost during the early stages results in a rapid expansion of its spreading range, exhibiting the most efficient growth effect among all indicators. These findings confirm that WCycle effectively balances cost and dissemination capability, making it a valuable tool for practical applications such as viral marketing and information diffusion strategies on social media platforms.

\section{Conclusion}\label{sec:conlusion}
Cycles are fundamental elements of network topology, capturing both structural and dynamic characteristics of complex systems. Despite the increasing use of cycle-based approaches in network analysis, their application in weighted networks remains underexplored, particularly with respect to integrating node behavioral traits. This paper addresses this gap by proposing Weighted Cycle (WCycle), a novel indicator that leverages basic cycles and edge weights to comprehensively assess node importance, incorporating both topological structure and behavioral information.

The effectiveness of WCycle has been validated through comparative analyses across multiple dimensions. First, WCycle demonstrates superior performance in identifying influential node groups, achieving consistently higher propagation capabilities compared to existing weighted centrality measures. It achieves this by selecting well-dispersed and structurally diverse key nodes, thereby mitigating influence redundancy and enhancing information dissemination. Second, structural and behavioral analyses reveal that WCycle effectively balances local and global node characteristics, enabling it to capture unique network features missed by traditional methods. Finally, the cost-performance evaluation highlights WCycle's efficiency, consistently achieving greater dissemination with lower initial costs, making it highly scalable and practical for real-world applications.

Nonetheless, this study has two limitations that suggest directions for future research. First, nodes that do not participate in any basic cycle are directly assigned an importance score of zero. This approach overlooks the potential significance of certain peripheral hub nodes, which may play crucial roles despite not being embedded in cycle structures. Second, the current method primarily focuses on identifying important node groups based on static structural information. However, a node or node group's influence is fundamentally determined by its dynamic behavioral attributes. Future research should investigate the feedback effects of cycles on network dynamics to capture a more comprehensive and accurate understanding of influence propagation, which is a more critical and challenging problem.

\section*{CRediT authorship contribution statement}
Wenxin Zheng: Conceptualization, Methodology, Software, Writing – original draft. Wenfeng Shi: Conceptualization, Methodology, Software, Validation. Tianlong Fan:  Conceptualization, Methodology, Validation, Writing - review \& editing, Supervision, Funding acquisition.  Linyuan Lü: Conceptualization, Writing – review \& editing, Supervision, Funding acquisition. 

\section*{Data and code availability}
The data and code supporting this study are available at \href{https://github.com/Wenxin02/WCycle}{https://github.com/Wenxin02/WCycle}.

\section*{Competing interests}
The authors declare no competing interests.

\bibliographystyle{IEEEtran}
\bibliography{IEEEabrv,main}

\newpage
\begin{IEEEbiography}[{\includegraphics[width=1in,height=1.25in,clip,keepaspectratio]{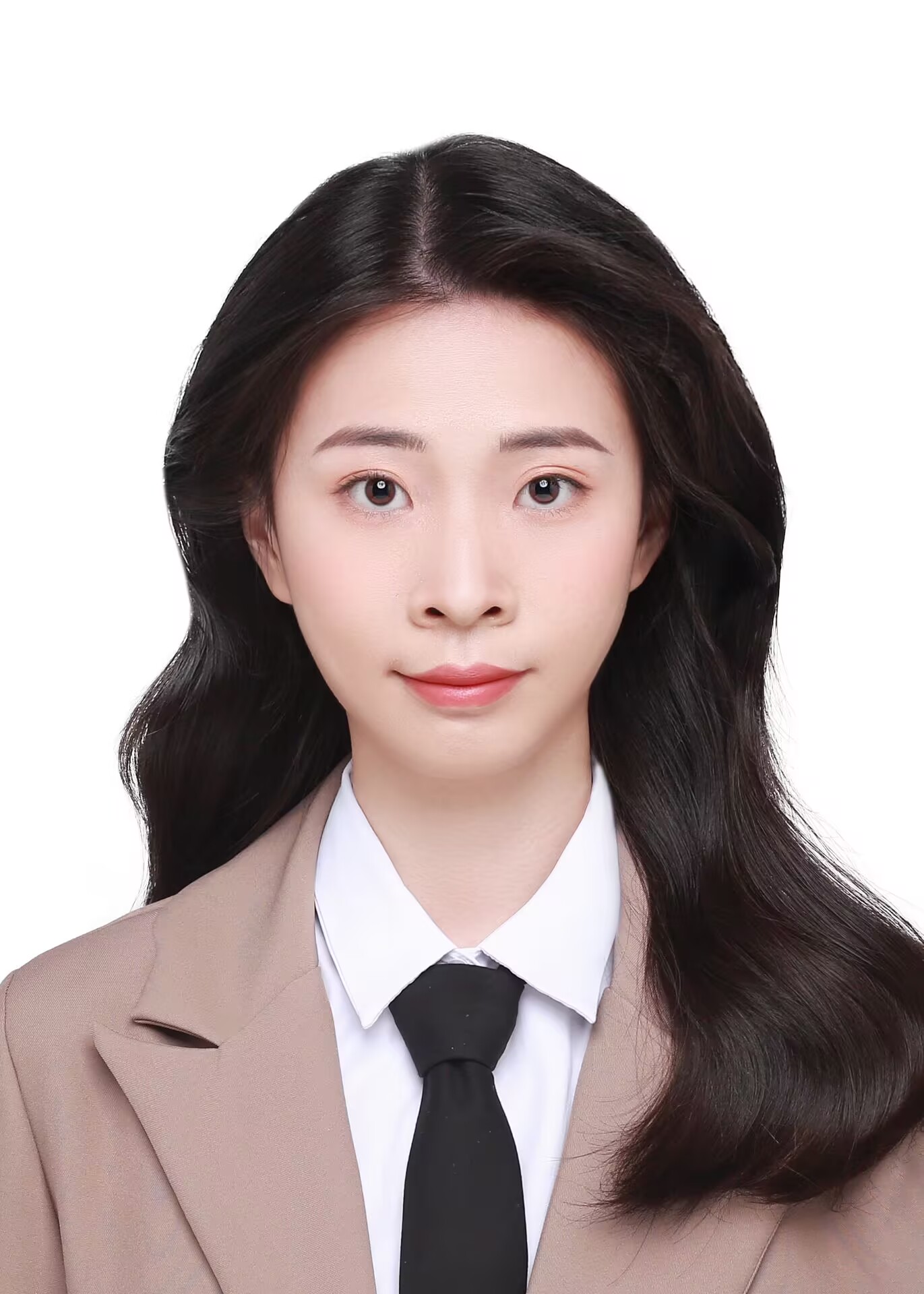}}]{Wenxin Zheng }
 received the B.Eng. degree in engineering from the University of Electronic Science and Technology of China, Chengdu, China, in 2024. She is currently pursuing the M.S. degree at the University of Science and Technology of China, Hefei, China. Her main research interests include the modeling and analysis of complex networks, with a focus on the identification of important nodes within complex networks.
\end{IEEEbiography}
\begin{IEEEbiography}[{\includegraphics[width=1in,height=1.25in,clip,keepaspectratio]{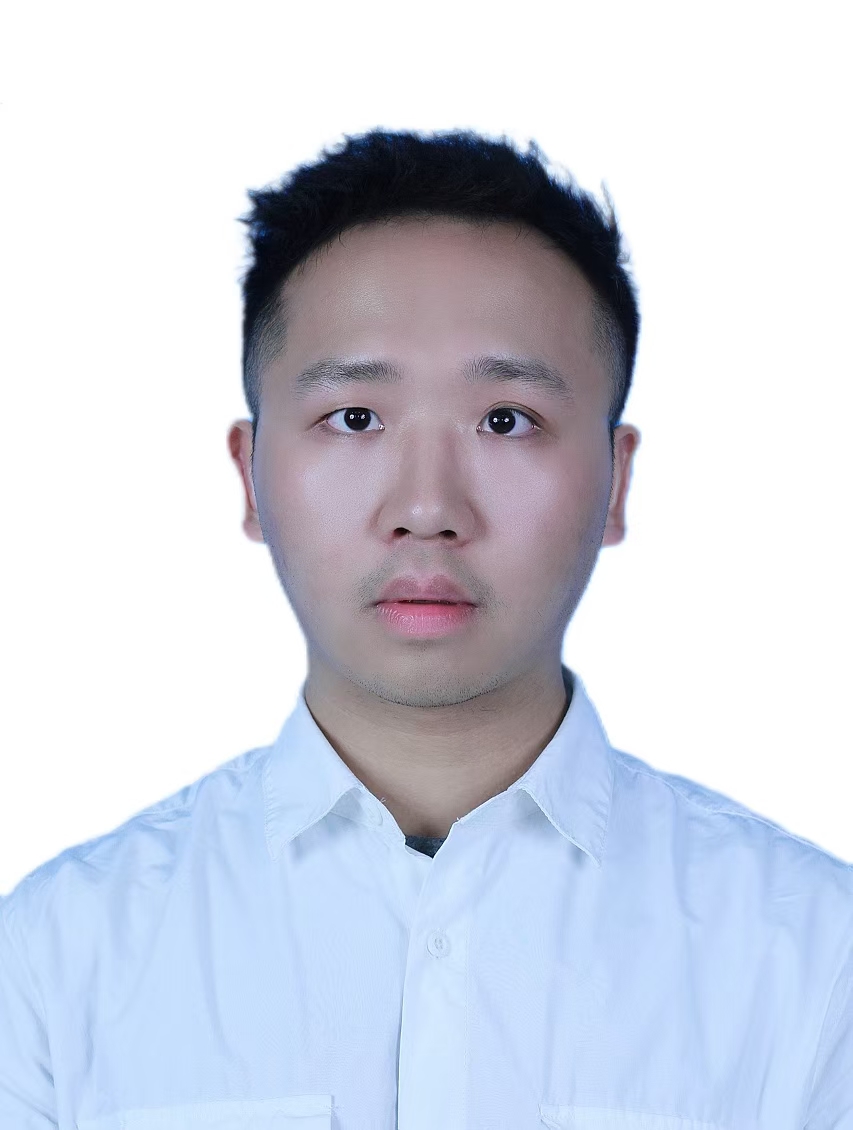}}]{Wenfeng Shi} received the MS degree in Computer Science from the University of Electronic Science and Technology of China, Chengdu, China, in 2023. He is currently working toward the PhD degree at University of Science and Technology of China. His main research interests include influence maximization and node ranking in complex networks.
\end{IEEEbiography}

\begin{IEEEbiography}[{\includegraphics[width=1in,height=1.25in,clip,keepaspectratio]{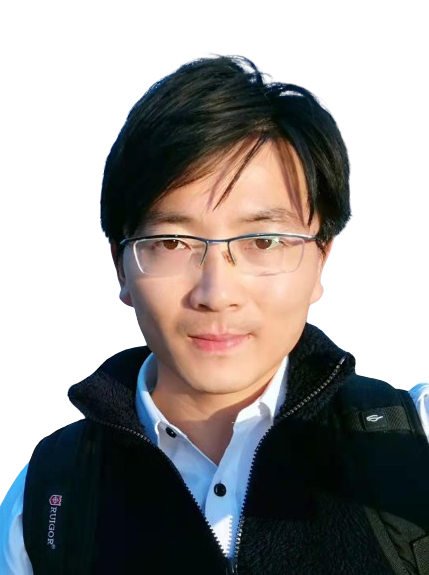}}]{Tianlong Fan} received the Ph.D. degree in theoretical interdisciplinary physics from the Universit\'e de Fribourg, Fribourg, Switzerland, in 2023. He is currently a postdoctoral researcher at the School of Cyber Science and Technology, University of Science and Technology of China. His current research interests include the theory and applications of complex networks and complex systems.
\end{IEEEbiography}

\begin{IEEEbiography}[{\includegraphics[width=1in,height=1.25in,clip,keepaspectratio]{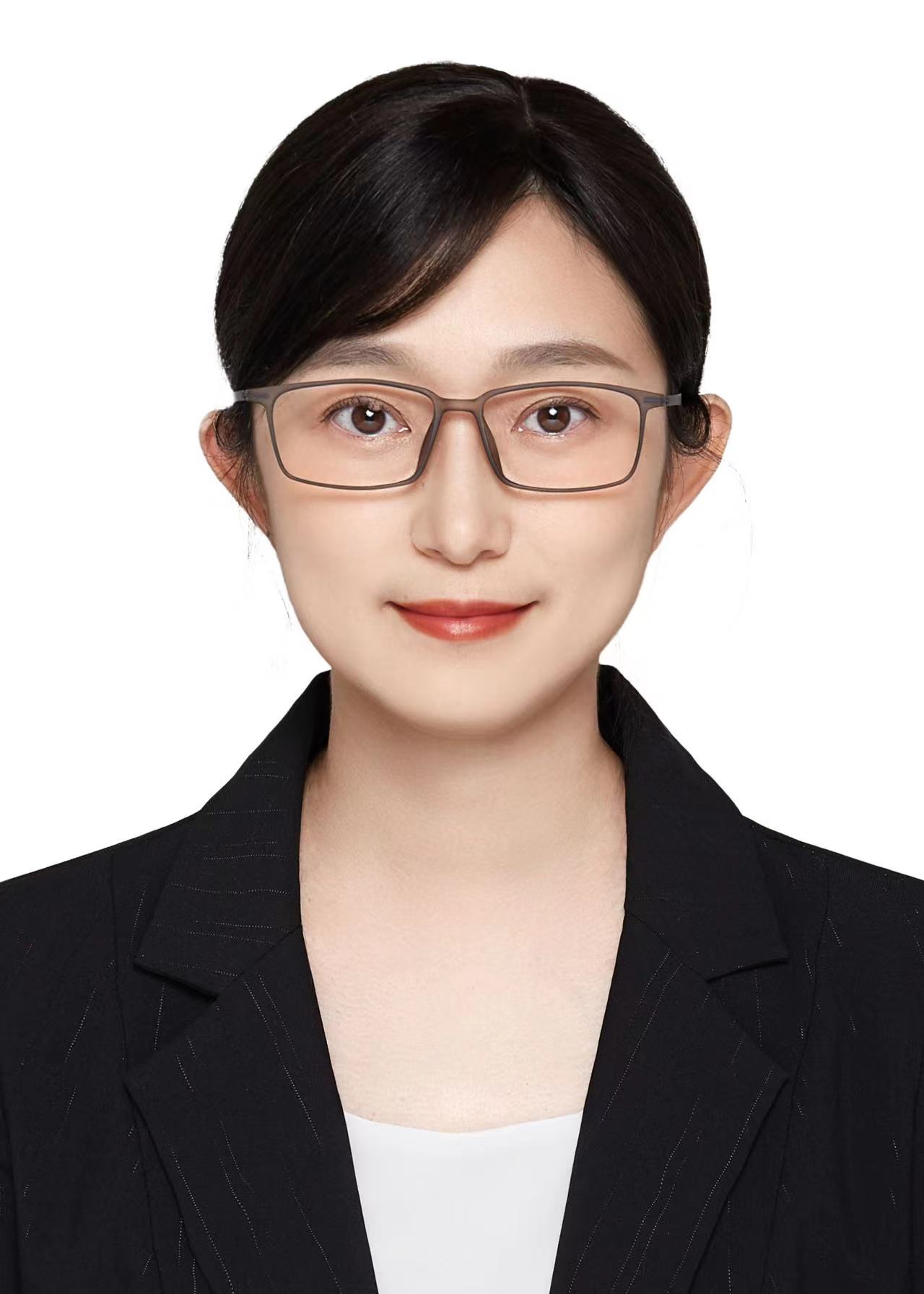}}]{Linyuan Lü} (Senior Member, IEEE) received the Ph.D. degree from the Universit\'e de Fribourg, Fribourg, Switzerland. She is currently a Professor at the School of Cyber Science and Technology, University of Science and Technology of China. She has utilized the theories and methods of statistical physics and network science to address a number of important issues in information sciences. Her current research interests include complex networks and social computing. She has published more than 80 research papers, of which the majority are published in prestigious peer-reviewed international journals, like Physics Reports, PNAS, Nature Communications, etc., including 9 ESI Top-1\% highly cited papers. She has applied for 21 invention patents and 10 of them were granted. Now she serves as a board member of the International Network Science Society.
\end{IEEEbiography}

\vfill

\end{document}